\documentclass[twocolumn,aps,pra]{revtex4-1}
\usepackage{amsmath}
\usepackage{amssymb}
\usepackage{graphicx}
\usepackage{esint}
\begin{document}

\title{Quantum Optics Model of Surface Enhanced Raman Spectroscopy for Arbitrarily
Shaped Plasmonic Resonators}

\author{Mohsen Kamandar Dezfouli and Stephen Hughes}

\date{\today}

\affiliation{Department of Physics, Engineering Physics and Astronomy, Queen's
University, Kingston, Ontario, Canada K7L 3N6}
\begin{abstract}
We present a self-consistent quantum optics approach to calculating
the surface enhanced Raman spectrum of molecules coupled to arbitrarily
shaped plasmonic systems. Our treatment is intuitive to use and provides
fresh analytical insight into the physics of the Raman scattering
near metallic surfaces and can be applied to a wide range of geometries
including resonators, waveguides, as well as hybrid photonic-plasmonic
systems. Our general theory demonstrates that the detected Raman spectrum
originates from an interplay between nonlinear light generation and
propagation (which also includes the effects of optical quenching).
Counter intuitively, at the nonlinear generation stage, we show that
the Stokes (anti-Stokes) signal at the molecule location depends on
the plasmonic enhancements, through the local density of photon states
(LDOS), at the anti-Stokes (Stokes) frequency. However, when propagating
from the vibrating molecule to the far field, the Stokes (anti-Stokes)
emission experiences a plasmonic enhancement at the Stokes (anti-Stokes)
frequency, as expected. We identify the limits of the commonly known
$E^{4}$ electric-field rule for Raman signal enhancement near plasmonic
surfaces at low pump powers, as well as a different $E^{8}$ rule
at high pump powers, and we clarify the role of the LDOS. Our theory
uses a generalized quantum master equation where the plasmonic system
is treated as an environmental bath that is described through the
photonic Green function of the medium. Therefore, any electrodynamics
related physics, such as quenching and propagation, are self-consistently
included in the model. The presented formalism is also capable of
spatially describing the full Raman response in a simple analytical
way, which is typically not possible using simpler coupled mode theories,
even if one adopts a quantum picture. This spatial analysis includes
both the dependency of the Raman signals at a fixed detector location
when the molecule is moved around the plasmonic platform, and the
dependency of the Raman signal on the detector location. We demonstrate
the power of this approach by using a quasinormal mode expansion theory
to construct the photonic Green functions of the plasmonic resonators,
and explore several different nanoresonator systems.
\end{abstract}
\maketitle

\section{Introduction}

\begin{figure}
\includegraphics[width=1\columnwidth]{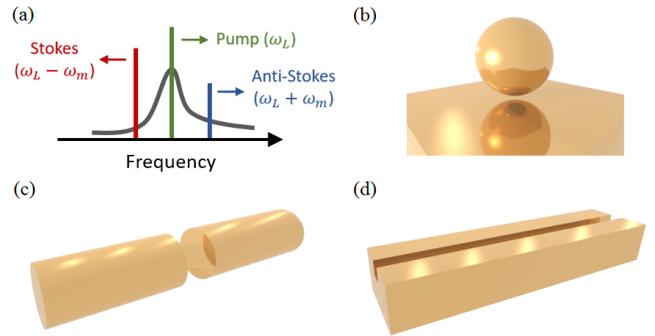}

\caption{(a). Picture of the energy levels involved in SERS where a molecule
of interest is coupled to a plasmonic environment. The broad, and
in general non-Lorentzian, plasmonic response is sketched where the
Stokes and anti-Stokes signal are enhanced according to strength of
the coupling regime they belong to. (b)-(d). Schematics of example
plasmonic structures that can be used to perform SERS.\label{fig:Schematic}}
\end{figure}

Plasmonic devices on the nano scale can be used to enhance the Raman
emission from molecules by many orders of magnitude \cite{Nie1997,Lal2007,Maier,SERS}.
This is greatly appreciated in sensing technologies, as the natural
Raman cross sections of different molecules are extremely small. The
significant enhancement comes from an ultratight confinement of light
that does not suffer from the diffraction limit. Even though the cavity
quality factor of typical plasmonic resonances are very low, perhaps
on the order of 10, the significant local fields for the plasmonic
resonances can lead to very large total spontaneous emission enhancement
factors, on the order of 1000 and more. Thus, the plasmonic devices
introduce new possibilities for next generation broadband chemical
sensors. In terms of design, a variety of different geometries can
be used for plasmonic light enhancement such as resonators \cite{ChemicalMapping,ResolvingE4,SeeingSingleMolecule,StrongCoupling},
waveguides \cite{Review2014,Wedge2015,Sensing2016}, and hybrid photonic-plasmonic
systems \cite{HybridSERS,Barth2010,Bassake2016} (see Fig.~\ref{fig:Schematic}
for some typical geometries of interest). Consequently, general theoretical
treatments of these devices must be flexible enough to be adopted
to a wide range of geometries, possibly with multiple resonances up
to even a continuum; while even at the single mode level (which has
some conceptual difficulties), a careful treatment of the normalized
fields used in a modal analysis is required (e.g., these are not simple
leaky cavity modes because of the material absorption).

In the past, several different approaches have been taken to theoretically
study Surface Enhanced Raman Spectroscopy (SERS) in plasmonic geometries,
where various ideas and objectives have been drawn \cite{prev1,prev3,DFT2008,prev2,john,DFT2014,prev4,SERS2016}.
For example, density functional theory (DFT) has been used to successfully
calculate the Raman spectra of molecules \cite{DFT2008,DFT2014}.
This approach can partly capture the effect of the surface roughness
of the nanoparticles on SERS, which is particularly important when
the molecule gets very close to the metal. On the other hand, if the
molecule is too near to the metallic surface (e.g., a few nanometer
away), extreme quenching at the metal surface becomes possible which
in turn leads to a poor quantum efficiency for photons to propagate
out radiatively and be detected. Moreover, the technical difficulty
of DFT schemes, along with the significant computational resources
required, can impose practical limitations in studying arbitrary geometries,
specially when the spatial extent of the device becomes prohibitively
large. More recently, a perturbation analysis of SERS has been proposed
where the plasmonic system is treated as a quasiparticle \cite{SERS2016}
and enhancements that are orders of magnitudes larger than what calculated
using standard electrodynamics theory, are predicted. While some of
the conclusions are somewhat controversial, e.g., the work suggests
that chemical enhancement can perhaps be ruled out in plasmonic SERS,
the key work is based on using the normal modes for the plasmonic
nanoparticles, which are known to be problematic, not only in lossy
plasmonic systems but in all open cavity systems in general \cite{Koenderink2010,Kristensen2012,KristensenACS,Sauvan2013}.

There has also been a semiclassical study of SERS where only the Raman
induced dipole is treated quantum mechanically \cite{john}. This
approach has been primarily developed for multilayer dielectric systems
and neglects the quantum mechanical interplay between the electromagnetic
field and the molecular vibrations, which can be especially important
at nonlinear higher pump powers. Such high field regimes are perhaps
more feasible using plasmonic systems as typically several orders
of magnitudes more enhancements are achievable compared to dielectric
systems. In addition to these works, there are recent quantum optomechanical
descriptions of SERS in plasmonic systems, where simple phenomenological
decay rates are used in a single mode description with Lorentzian
line shapes \cite{OptomechanicsRaman,QEDpaperRaman}. While these
approaches are intuitive, not too complex, and can be in principle
be implemented for simple cavity geometries, the computational cost
for extending beyond a single mode picture can be prohibitive, arising
from the fact that very large Hilbert spaces are commonly needed for
converged results. However, if the complexity of having multiple overlapping
resonances can be dealt with, such a quantum mechanical picture can
lead to a computationally cost-effective model for study of SERS near
plasmonic systems with arbitrary geometry, where one can go well beyond
an approximate single mode picture.

In this work, we use an open system quantum optics approach to derive
analytical expressions for the spatially-dependent detected Stokes
and anti-Stokes intensities for molecules coupled to a general plasmonic
system. Our results can be directly applied in their current form
to arbitrarily shaped geometries (including dielectric systems of
course), with no additional computational cost than calculating the
system electromagnetic response which must be done in any case to
understand the modal properties of the medium. We use a generalized
master equation \cite{Carmichael,Breuer,GME} such that the exact
system photon Green function is exploited in the formalism, which
rigorously includes the full frequency response of the plasmonic enhancement
and quenching effects; this avoids having to adopt an approximate
coupled mode theory with dissipation, which is usually used in many
cavity-QED formalisms, including plasmonic cavity systems.

We start with a representation of the electromagnetic Hamiltonian
in the basis of a continuum of quantum field operators, valid for
any inhomogeneous lossy medium \cite{Welsch,Suttorp}. Then, the full
system dynamics is projected onto the basis of the molecule vibrational
mode by tracing out the plasmonic degrees of freedom, treating the
plasmonic system as a bath or reservoir, but one that is treated in
a self-consistent way. This is expected to be valid as long as the
molecule is not strongly coupled to the plasmonic system. Since the
molecule vibration or phonon mode is described by a single mode harmonic
oscillator, the resulting master equation is of a very simple form
that can be solved analytically for different observables of interest,
fully including the plasmonic reservoir. In particular, from the obtained
master equation, we derive the spectrum analytically, and connect
to approximate SERS theories in the literature. In addition, we use
our analytical expressions for both Stokes and anti-Stokes intensities
to compute a spatial map of the Raman spectrum, either when the molecule
is moved nearby the plasmonic system, or when the detector location
is scanned. In our main analysis, we consider a gold dimer resonator
made of two cylindrical nanorods as shown in Fig.~\ref{fig:Schematic}.(c)
as an example system. However, we stress that no modification to the
existing theory is needed for different cavity structures, and one
simply requires the Green function for the new system. In order to
show such potential for studying arbitrary metal structures, we also
consider a hybrid photonic-plasmonic system such that the already
mentioned dimer is now placed on top of a photonic crystal nanobeam
cavity. This is certainly beyond the usual single mode picture for
the photonic medium and demonstrates the generality of our approach.

With regards to SERS, the total Raman signal is known to be partially
dependent on the field enhancement at the laser frequency, $\omega_{L}$,
as well as field enhancement at the Raman frequency, $\omega_{R}$,
through \cite{SERS} 
\begin{equation}
{\rm EF}\left(\omega_{R},\omega_{L}\right)\propto\dfrac{\left|E\left(\omega_{R}\right)\right|^{2}\left|E\left(\omega_{L}\right)\right|^{2}}{\left|E_{0}\left(\omega_{R}\right)\right|^{2}\left|E_{0}\left(\omega_{L}\right)\right|^{2}},\label{eq:EF}
\end{equation}
which is most often stated as the $E^{4}$ electric-field law for
total enhancement, especially when laser and Raman frequencies are
close. While our model recovers this general enhancement factor for
the detected Raman intensities, our findings shed new light on the
details of the Raman process, particularly differentiating between
the signal generating at the molecule location and photon detection
on the detectors placed in the far-field. We also show that the $E^{4}$
enhancement factor comes from propagation effects at Raman frequencies
as well as the LDOS enhancement at the pump frequency, but the Stokes
generation is sensitive to the plasmonic local density of states (LDOS)
at anti-Stokes frequency and vice-versa, a non-trivial notion that
to the best of our knowledge has not been reported before in the literature.
In addition, we identify a second (nonlinear) enhancement regime depending
on the strength of the pump field. We argue that, for high enough
pump intensities, the induced Raman emission enhancement factor can
follow a $E^{8}$ electric-field rule instead.

The rest of our paper is organized as follows. In Sec.~\ref{sec:Theory},
we introduce the details of our SERS theory, including the quantized
system Hamiltonian in arbitrary dispersive and lossy media; we also
present our generalized master equation projected onto the basis of
the molecule, and derive the spatially dependent spectrum calculated
using the expectation values of two-time correlation functions; in
addition, for a localized plasmon resonance, we use the approach of
quasinormal modes (QNMs) as a basis to accurately obtain the plasmonic
Green function, allowing us to study the spatial dependence of SERS
in a remarkably clear and simply way. In the same section, we also
briefly discuss the key elements of an alternative quantum optomechanical
model using coupled modes for the resonator and a molecule, but restricted
to a single plasmon mode picture~\cite{QEDpaperRaman}; we later
use this for comparison and verification of our general model in this
limit. Next, in Sec.~\ref{sec:Results}, we adopt our analytical
expression for the Raman spectrum to perform a detailed study of SERS
using the two example plasmonic systems mentioned above. Finally,
we present our conclusions in Sec.~\ref{sec:Conclusions}.

\section{Theory\label{sec:Theory}}

In this section, we present the key elements of our formalism including
the system Hamiltonian in quantized form and the generalized master
equation which must be solved for the optical spectrum to be obtained.
We also briefly discuss the QNM representation of the system Green
function that can be used to accurately estimate the emission characteristics
of two different plasmonic cavity systems that we study.

\subsection{System Hamiltonian}

The molecular vibration of interest can be considered as a simple
Harmonic oscillator, where the corresponding Hamiltonian can be written
as 
\begin{equation}
H_{{\rm m}}=\hbar\omega_{m}b^{\dagger}b,
\end{equation}
where $b$ and $b^{\dagger}$ satisfy the Bosonic commutation relation
$\left[b,b^{\dagger}\right]=1$ and $\omega_{m}$ is the real-valued
frequency of the vibrational mode. In reality, a single molecule supports
more than one vibrational mode at different frequencies, however in
practice, usually one of them in particular is under investigation.
Quite generally, the free field Hamiltonian in a lossy medium can
be expressed in terms of a continuum of field operators, $\mathbf{f}\left(\mathbf{r},\omega\right)$,
such that \cite{Welsch,Suttorp} 
\begin{align}
H_{{\rm ph}} & =\hbar\int d{\bf r}\int_{0}^{\infty}d\omega\,\omega\,\mathbf{f}^{\dagger}\left(\mathbf{r},\omega\right)\mathbf{f}\left(\mathbf{r},\omega\right),
\end{align}
where $\mathbf{f}$ and $\mathbf{f}^{\dagger}$ follow Bosonic algebra
with commutation rules, 
\begin{equation}
\left[\mathbf{f}\left(\mathbf{r},\omega\right),\mathbf{f}^{\dagger}\left(\mathbf{r}^{\prime},\omega^{\prime}\right)\right]=\mathbf{I}\,\delta\left(\mathbf{r}-\mathbf{r}^{\prime}\right)\delta\left(\omega-\omega^{\prime}\right).
\end{equation}
Importantly, our picture of the photonic environment is not restricted
to a single mode case. As mentioned in the introduction, this is necessary
in general, as many plasmonic environments do not satisfy a single
mode regime. For example, in case of plasmonic waveguides, naturally
there is a continuum of propagating modes involved. In the case of
hybrid photonic-plasmonic systems, there are at very least two modes
involved. This later case is also true for many plasmonic resonators,
especially when small gaps sizes are used, such as the one schematically
shown in Fig.~\ref{fig:Schematic}(b). Even in the single mode regime,
plasmonic resonances are described using QNMs that are solutions to
a non-Hermitian wave equation, and subsequently the quantization procedure
is non trivial. Thus, it is not obvious how one should include the
important medium parameters, such as the LDOS, propagation of light
from the molecule to detector, as well as quenching in the quantum
mechanical theory of SERS. The usefulness of the continuum picture
used above becomes further revealed when the electric-field operator
for a lossy medium described by the complex permittivity function
$\varepsilon\left(\mathbf{r},\omega\right)$ is expanded in terms
of the field operators, $\mathbf{f}\left(\mathbf{r},\omega\right)$,
using \cite{Welsch,Suttorp} 
\begin{align}
 & \mathbf{E}\left(\mathbf{r},\omega\right)=\nonumber \\
 & \ \ \frac{1}{\varepsilon_{0}}\int d\mathbf{r}^{\prime}\mathbf{G}\left(\mathbf{r},\mathbf{r}^{\prime};\omega\right)\cdot\sqrt{\frac{\hbar\varepsilon_{0}}{\pi}{\rm Im\left\{ \varepsilon\left(\mathbf{r}^{\prime},\omega\right)\right\} }}\mathbf{f}\left(\mathbf{r}^{\prime},\omega\right),\label{eq:eInlossy}
\end{align}
where the two space point classical Green function, $\mathbf{G}\left(\mathbf{r},\mathbf{r}^{\prime};\omega\right)$
is the solution to the classical Helmholtz equation 
\begin{align}
 & \nabla\times\nabla\times\mathbf{G}\left(\mathbf{r},\mathbf{r}^{\prime};\omega\right)-\frac{\omega^{2}}{c^{2}}\varepsilon\left(\mathbf{r},\omega\right)\mathbf{G}\left(\mathbf{r},\mathbf{r}^{\prime};\omega\right)\nonumber \\
 & =\frac{\omega^{2}}{c^{2}}\mathbf{I}\,\delta\left(\mathbf{r}-\mathbf{r}^{\prime}\right),\label{eq:GFdefine}
\end{align}
subjected to, in general, open boundary condition. Note that our defining
equation for the Green function is slightly different than another
commonly used notation \cite{nanoOptics} in its right hand side.
When we derive our master equation below, the plasmonic degrees of
freedom will be traced out and effectively the classical Green function,
$\mathbf{G}\left(\mathbf{r},\mathbf{r}^{\prime};\omega\right)$, remains
to characterize the plasmonic behavior in our model. This reservoir
is justified as the molecule is typically weakly coupled to the plasmonic
system (but in the Purcell regime), especially with their low quality
factors.

In a dipole interaction, the induced Raman polarization, $\mathbf{p}_{R}$,
can be written as 
\begin{equation}
\mathbf{p}_{R}=\boldsymbol{\alpha}_{R}\cdot\mathbf{E}\left(\mathbf{r}_{m}\right),
\end{equation}
where $\boldsymbol{\alpha}_{R}$ is the Raman polarizability tensor
of the molecule that can be expressed in terms of the well-known Raman
tensor of the molecule, $\mathbf{R}$, and the quantized displacement
via \cite{SERS} 
\begin{equation}
\boldsymbol{\alpha}_{R}=\mathbf{R}\,\sqrt{\frac{\hbar}{2\omega_{m}}}\left(b+b^{\dagger}\right),
\end{equation}
and $\mathbf{E}\left(\mathbf{r}_{m}\right)$ is the total electric
field operator at the molecule location $\mathbf{r}_{m}$, defined
as 
\begin{equation}
\mathbf{E}\left(\mathbf{r}\right)=\int d\omega\mathbf{E}\left(\mathbf{r},\omega\right)+{\rm H.c.}
\end{equation}
Thus, the interaction Hamiltonian associated with the Raman induced
dipole oscillation is 
\begin{align}
H_{I} & =-\mathbf{p}_{R}\cdot\mathbf{E}\left(\mathbf{r}_{m}\right)\\
 & =-\sqrt{\frac{\hbar}{2\omega_{m}}}\left(b+b^{\dagger}\right)\mathbf{E}\left(\mathbf{r}_{m}\right)\cdot\mathbf{R}\cdot\mathbf{E}\left(\mathbf{r}_{m}\right).\label{eq:IntHam}
\end{align}
This form of the system Hamiltonian is similar to the (coupled-mode)
optomechanical Hamiltonian used in \cite{OptomechanicsRaman,QEDpaperRaman},
which we discuss in more details in subsection~\ref{subsec:Existing-quantum-optomechanical};
however, here we use a continuum representation of the electromagnetic
field, and the classical Green function of the system is incorporated
in a self-consistent manner through use of Eq.~\eqref{eq:eInlossy}.

\subsection{Generalized master equation}

We now use the above Hamiltonian to show how a generalized master
equation can be constructed, when the plasmonic system is treated
as a photonic reservoir. To proceed, we note that the Raman scattering
mechanism is a nonlinear optical process that the pump filed at frequency
$\omega_{L}$ induces a polarization that oscillates at the Raman
frequency, $\omega_{R}=\omega_{L}\pm\omega_{m}$. Therefore, it is
natural to assume that the $\mathbf{E}\left(\mathbf{r}_{m}\right)$
on the left side of the Raman tensor in Eq.~\eqref{eq:IntHam} includes
only the pump frequency in its expansion, while the $\mathbf{E}\left(\mathbf{r}_{m}\right)$
on the right side includes all other frequencies but the pump. In
addition, we use a classical representation of the pump field to simplify
the mathematics involved, which is valid for the high pump fields
of interest; thus for the pump field, we are working with averaged
field operators where small quantum mechanical fluctuations are neglected
(which is consistent with the fact that the field operators are going
to be traced out eventually).

For the initial pump field, we consider a continuous wave (CW) excitation
with amplitude $\mathbf{F}_{0}$ at frequency $\omega_{L}$. In the
presence of the scattering geometry, the pump field at the molecule
location is 
\begin{equation}
\mathbf{F}_{{\rm p}}\left(\mathbf{r}_{m},t\right)=\eta\left[\mathbf{F}_{0}\left(\mathbf{r}_{m}\right)e^{i\omega_{L}t}+\mathbf{F}_{{\rm 0}}^{*}\left(\mathbf{r}_{m}\right)e^{-i\omega_{L}t}\right],
\end{equation}
where $\eta$ is the external field enhancement factor and can be
calculated through use of the system Green function: 
\begin{equation}
\eta=1+\dfrac{\int\left[\varepsilon\left(\mathbf{r},\omega_{L}\right)-\varepsilon_{B}\right]\mathbf{n}\cdot\mathbf{G}\left(\mathbf{r}_{m},\mathbf{r};\omega_{L}\right)\cdot\mathbf{F}_{0}\left(\mathbf{r}\right)d{\bf r}}{\mathbf{n}\cdot\mathbf{F}_{0}\left(\mathbf{r}_{m}\right)}.\label{eq:eta}
\end{equation}
Here, the unit vector $\mathbf{n}$ represents the direction of the
induced Raman dipole which is determined by the dominant Raman tensor
element, $R_{nn}$. Note that the value of $\eta$ implicitly depends
on the molecule location, $\mathbf{r}_{m}$, and the laser frequency,
$\omega_{L}$. The dielectric function difference involved in this
expression, where $\varepsilon_{B}$ refers to the background medium,
implies that the spatial integration takes place only over the volume
of the metallic nanoparticles. Also, note that, we have used $\mathbf{F}$
for the classical pump field to reserve $\mathbf{E}$ as a quantum
mechanical electric field operator, and ${\bf F}_{{\rm p}}$ can be
considered as $\left\langle \mathbf{E}_{{\rm p}}\right\rangle $.

The general form of the time local master equation in the interaction
picture (tilde represents the interaction picture), using a Born-Markov
approximation is \cite{Carmichael,Breuer} 
\begin{align}
 & \frac{\partial\tilde{\rho}\left(t\right)}{\partial t}=\nonumber \\
 & \ \ \ -\frac{1}{\hbar^{2}}\int_{0}^{t}d\tau\,{\rm tr_{R}}\left\{ \left[\tilde{H}_{I}\left(t\right),\left[\tilde{H}_{I}\left(t-\tau\right),\tilde{\rho}\left(t\right)\rho_{{\rm R}}\right]\right]\right\} ,\label{eq:ME}
\end{align}
where $\tilde{\rho}$ is the reduced density matrix in the basis of
molecular vibrations, $\rho_{{\rm R}}$ is the plasmonic reservoir
density that the trace will be performed over, and $\tilde{H}_{I}$
is the interaction Hamiltonian in the interaction picture. Working
at optical range of frequencies allows one to use the following bath
approximation for the field operators, i.e., 
\begin{align}
{\rm tr_{R}\left\{ \mathbf{f}^{\dagger}\left(\mathbf{r},\omega\right)\mathbf{f}\left(\mathbf{r}^{\prime},\omega^{\prime}\right)\rho_{R}\right\} } & =0\\
{\rm tr_{R}\left\{ \mathbf{f}\left(\mathbf{r},\omega\right)\mathbf{f}^{\dagger}\left(\mathbf{r}^{\prime},\omega^{\prime}\right)\rho_{R}\right\} } & =\mathbf{I}\,\delta\left(\mathbf{r}-\mathbf{r}^{\prime}\right)\delta\left(\omega-\omega^{\prime}\right).
\end{align}
Transforming the Hamiltonian \eqref{eq:IntHam} to the interaction
picture, then substituting into the master equation \eqref{eq:ME},
and using the above bath approximations, we obtain the reduced density
matrix of the system in the basis of the molecule, 
\begin{align}
\frac{\partial\tilde{\rho}}{\partial t} & =J_{{\rm ph}}\left(\omega_{L}+\omega_{m}\right)\left(2b\tilde{\rho}b^{\dagger}-b^{\dagger}b\tilde{\rho}-\tilde{\rho}b^{\dagger}b\right)\nonumber \\
 & +J_{{\rm ph}}\left(\omega_{L}-\omega_{m}\right)\left(2b^{\dagger}\tilde{\rho}b-bb^{\dagger}\tilde{\rho}-\tilde{\rho}bb^{\dagger}\right),\label{eq:GME_int}
\end{align}
where $J_{{\rm ph}}$ represents the strength of the plasmonic induced
Raman emission. As seen, $J_{{\rm ph}}$ depends on the dominant Raman
tensor element, $R_{nn}$, the field enhancement factor, $\eta$,
and\textemdash most importantly\textemdash the plasmonic LDOS through
\begin{equation}
J_{{\rm ph}}\left(\omega\right)\equiv\frac{R_{nn}^{2}\left|\eta\right|^{2}\left|\mathbf{n}\cdot\mathbf{F}_{0}\right|^{2}}{2\varepsilon_{0}\omega_{m}}\,{\rm Im}\left\{ G_{nn}\left(\mathbf{r}_{{\rm m}},\mathbf{r}_{{\rm m}};\omega\right)\right\} .\label{eq:defineJph}
\end{equation}
Note that, in going from Eq.~\eqref{eq:GME} to Eq.~\eqref{eq:GME_int},
the time integral upper limit has been extended to infinity, as the
time scale for system changes are assumed to be much longer than its
counterpart for the reservoir.

In Eq.~\eqref{eq:GME_int}, the first term represents the Stokes
signal (which recall is emitted at $\omega_{L}-\omega_{m}$) proportional
to $J_{{\rm ph}}\left(\omega_{L}+\omega_{m}\right)$, and the second
term represent the anti-Stokes signal (emitted at $\omega_{L}+\omega_{m}$)
proportional to $J_{{\rm ph}}\left(\omega_{L}-\omega_{m}\right)$.
This means that, at the driving/generation stage, the Stokes emission
is governed by the plasmonic enhancements via the projected LDOS at
the anti-Stokes frequency and vice-versa; this may seem contrary to
the common notion that SERS involves field enhancement at the corresponding
frequencies for both Stokes and anti-Stokes signals, and moreover
it identifies a different operating regime that can be achieved at
high pump powers. As will be discussed later, the commonly seen enhancement
factor of Eq.~\eqref{eq:EF} is a consequence of the propagation
as well as pump field enhancement in the presence of the scattering
geometry, particularly when low pump powers are considered. It is
also worth mentioning that such features are not totally surprising
in the domain of quantum nonlinear optics, e.g., such a non-trivial
LDOS coupling has also been noted for the Mollow triplet spectrum
from excited quantum dots coupled to plasmonic resonators, where the
center peak width depends only on the LDOS at the two side peaks and
not at all on the LDOS at the laser drive frequency \cite{GePRB2013}.

To obtain the full master equation of the system, the coupling of
the Raman vibrations to the the thermal environmental bath must be
also accounted for. This can be done using additional Lindblad terms
associated with thermal dissipation as well as incoherent pumping
\cite{QEDpaperRaman}. Thus, one finally arrives at the following
master equation for the complete molecule-plasmonic system: 
\begin{align}
\frac{\partial\rho}{\partial t} & =-i\omega_{m}\left[b^{\dagger}b,\rho\right]\nonumber \\
 & +J_{{\rm ph}}\left(\omega_{L}+\omega_{m}\right)\left(2b\rho b^{\dagger}-b^{\dagger}b\rho-\rho b^{\dagger}b\right)\nonumber \\
 & +J_{{\rm ph}}\left(\omega_{L}-\omega_{m}\right)\left(2b^{\dagger}\rho b-bb^{\dagger}\rho-\rho bb^{\dagger}\right)\nonumber \\
 & +\gamma_{m}\left(\bar{n}^{{\rm th}}+1\right)\left(2b\rho b^{\dagger}-b^{\dagger}b\rho-\rho b^{\dagger}b\right)\nonumber \\
 & +\gamma_{m}\bar{n}^{{\rm th}}\left(2b^{\dagger}\rho b-bb^{\dagger}\rho-\rho bb^{\dagger}\right),\label{eq:GME}
\end{align}
where the overhead tilde is removed now, as we have moved back to
the Schrödinger picture. This generalized master equation can be used
to evaluate the time dependence of different operators of interest.
In particular, to connect to the Raman detection experiments, this
is used in next to obtain an analytical expression for the spatially
dependent spectrum.

\subsection{Detected spectrum on a point detector\label{subsec:Detected-spectrum-on}}

The detectable spectrum at $\mathbf{R}$ is defined as 
\begin{equation}
S\left(\mathbf{R},\omega\right)\equiv\left\langle \mathbf{E}^{\dagger}\left(\mathbf{R},\omega\right)\cdot\mathbf{E}\left(\mathbf{R},\omega\right)\right\rangle ,\label{eq:spectrum1}
\end{equation}
where we assume a point detector. The molecule polarization emission
transfers to this point (typically in the far field, but not necessarily
so) through the system Green function using \cite{Yao2009} 
\begin{equation}
\mathbf{E}\left(\mathbf{R},\omega\right)=\frac{1}{\varepsilon_{0}}\mathbf{G}\left(\mathbf{R},\mathbf{r}_{{\rm m}};\omega\right)\cdot\mathbf{p}_{R}\left(\mathbf{r}_{{\rm m}};\omega\right),
\end{equation}
which is an exact input-output expression that is derived for the
Heisenberg equations of motion. Therefore, using the field operator
and the molecular polarization definitions, we find 
\begin{equation}
S\left(\mathbf{R},\omega\right)=\frac{\hbar R_{nn}^{2}\left|\eta\right|^{2}\left|\mathbf{n}\cdot\mathbf{F}_{0}\right|^{2}}{2\omega_{m}\varepsilon_{0}^{2}}\left|\mathbf{G}\left(\mathbf{R},\mathbf{r}_{{\rm m}};\omega\right)\cdot\mathbf{n}\right|^{2}S_{0}\left(\omega\right),\label{eq:spectrum2}
\end{equation}
where $S_{0}\left(\omega\right)=S_{0}^{{\rm st}}\left(\omega\right)+S_{0}^{{\rm as}}\left(\omega\right)$
is the spectrum calculated using the molecule dynamics that is a sum
of Stokes, $S_{0}^{{\rm st}}\left(\omega\right)$, and anti-Stokes,
$S_{0}^{{\rm as}}\left(\omega\right)$, signals, defined from 
\begin{align}
S_{0}^{{\rm st}}\left(\omega\right) & \equiv\left\langle b\left(\omega\right)b^{\dagger}\left(\omega\right)\right\rangle \\
 & ={\rm Re}\left\{ \int_{0}^{\infty}d\tau e^{i\left(\omega-\omega_{L}\right)\tau}\left\langle b\left(t\right)b^{\dagger}\left(t+\tau\right)\right\rangle \right\} ,
\end{align}
and 
\begin{align}
S_{0}^{{\rm as}}\left(\omega\right) & \equiv\left\langle b^{\dagger}\left(\omega\right)b\left(\omega\right)\right\rangle \\
 & ={\rm Re}\left\{ \int_{0}^{\infty}d\tau e^{i\left(\omega-\omega_{L}\right)\tau}\left\langle b^{\dagger}\left(t\right)b\left(t+\tau\right)\right\rangle \right\} .
\end{align}
Here, we have excluded $\left\langle b\left(\omega\right)b\left(\omega\right)\right\rangle $
and $\left\langle b^{\dagger}\left(\omega\right)b^{\dagger}\left(\omega\right)\right\rangle $
terms as they are associated with higher order Raman intensities.
Using the quantum regression theorem \cite{Carmichael}, equations
of motion for the two-time expectation values can be derived from
the generalized master equation of Eq.~\eqref{eq:GME}, which in
turn can be solved analytically using Laplace transform techniques
to arrive at 
\begin{align}
 & S_{0}^{{\rm st}}\left(\omega\right)=\nonumber \\
 & {\rm Re}\left\{ \dfrac{i\left[\gamma_{m}\left(\bar{n}^{{\rm th}}+1\right)+J_{{\rm ph}}\left(\omega_{L}+\omega_{m}\right)\right]}{\left[\omega-\left(\omega_{L}-\omega_{m}\right)+i\left(\gamma_{m}+\Delta J_{{\rm ph}}\right)\right]\left(\gamma_{m}+\Delta J_{{\rm ph}}\right)}\right\} ,\label{eq:spectrum3}\\
 & S_{0}^{{\rm as}}\left(\omega\right)=\nonumber \\
 & {\rm Re}\left\{ \dfrac{i\left[\gamma_{m}\bar{n}^{{\rm th}}+J_{{\rm ph}}\left(\omega_{L}-\omega_{m}\right)\right]}{\left[\omega-\left(\omega_{L}+\omega_{m}\right)+i\left(\gamma_{m}+\Delta J_{{\rm ph}}\right)\right]\left(\gamma_{m}+\Delta J_{{\rm ph}}\right)}\right\} ,\label{eq:spectrum4}
\end{align}
where for convenience we have defined $\Delta J_{{\rm ph}}=J_{{\rm ph}}\left(\omega_{L}+\omega_{m}\right)-J_{{\rm ph}}\left(\omega_{L}-\omega_{m}\right)$.
Therefore, we have succeeded in deriving a transparent analytical
solutions to the SERS spectrum for any plasmonic system, provided
by Eqs.~\eqref{eq:spectrum2}, \eqref{eq:spectrum3} and \eqref{eq:spectrum4};
by including the classical system Green function for the electromagnetic
field, this approach brings considerable power to the existing theory
compared to the previous theories.

Before presenting some representative calculations, we first discuss
some of the aspects of the analytical SERS spectrum obtained using
our generalized Master equation. We begin with the effect of increasing
the pump power on the Raman intensities. Let us consider the Stokes
emission given in Eq.~\eqref{eq:spectrum3} for two different cases:
(i) low pump limit where the thermal pumping quantified by $\gamma_{m}\left(n^{{\rm th}}+1\right)$
is dominant over plasmonic drive $J_{{\rm ph}}\left(\omega_{L}+\omega_{m}\right)$;
and (ii) high pump limit where plasmonic induced emission become comparable
to or stronger than thermal phonons. Similar conclusions hold for
the anti-Stokes signal given in Eq.~\eqref{eq:spectrum4}. In the
former scenario, the counter-intuitive enhancement process explained
earlier remains unseen, and ignored in most works on SERS, especially
because the very low values for the Raman tensor elements lead to
small $J_{{\rm ph}}$ values. In such cases, the maximum Stokes generation
is approximately 
\begin{equation}
S_{0}^{{\rm st}}\left(\omega_{L}-\omega_{m}\right)\approxeq\dfrac{\bar{n}^{{\rm th}}+1}{\gamma_{m}},
\end{equation}
where essentially the Stokes generation becomes weakly dependent on
the plasmonic enhancement. On the other hand, with the strong modal
enhancements possible in plasmonic systems, sufficiently large values
for the quantity $J_{{\rm ph}}$ can be obtained such that the Stokes
generation is mainly derived by the plasmonic enhancement rather than
thermal vibrations. Assuming that $\Delta J_{{\rm ph}}$ is small
compared to $\gamma_{m}$, then the maximum Stokes generation is now
approximately 
\begin{equation}
S_{0}^{{\rm st}}\left(\omega_{L}-\omega_{m}\right)\approxeq\dfrac{J_{{\rm ph}}\left(\omega_{L}+\omega_{m}\right)}{\gamma_{m}^{2}}.
\end{equation}
This is a regime where there is a strong interplay between the plasmonic
resonator and the molecular vibrations. The plasmonic enhancement
enters in a much stronger fashion through use of Eq.~\eqref{eq:defineJph}.
Such nonlinear interplay between the plasmonic resonator and the Raman
vibrations has been only recently numerically predicted to be responsible
for blue-shifting the peak of the anti-Stokes intensity at high pump
powers \cite{QEDpaperRaman}. Here, we explain the physics behind
such an interplay.

Next, we emphasize that the classical system Green function basically
enters our formalism twice; once, depending on the pump strength,
possibly at the generation stage where Stokes and anti-Stokes emissions
are influenced by the dressed-state sampling the plasmonic resonance
at their co-frequencies, and then at the propagation stage where the
generated Stokes and anti-Stokes photons propagate to the far-field
through the plasmonic resonance at their corresponding frequencies.
This later effect, along with the effective field enhancement at laser
frequency enforced by $\eta$, recovers the well-known $E^{4}$ enhancement
rule for SERS at low pump powers and leads to the much stronger $E^{8}$
enhancement rule at high pump power. This is hidden in our notation
using the system Green function, but will become more transparent
in subsection \ref{subsec:QNM-expansion-of}.

In addition, because the propagation effects are self-consistently
included in the definition of the system Green function (including
any quenching effects), there is no need for phenomenological effects
to be added by hand. For example, the $\omega^{4}$ dependence for
the detected spectrum considered in \cite{QEDpaperRaman} comes from
a simple dipole emission power argument, whereas in approach it is
automatically captured and accounted for in our formulation. Along
the same lines, our theory includes the location information of the
molecule as well as the detector. Therefore, Stokes and anti-Stokes
detectable intensities can be both mapped out when either the molecule
or the detector position is changed.

\subsection{ Coupled-mode quantum optomechanical model\label{subsec:Existing-quantum-optomechanical}}

In this subsection, we briefly present the key elements of the normal
mode quantum optomechanical studies of SERS, elegantly introduced
in Refs.~\cite{OptomechanicsRaman,QEDpaperRaman}, to help the reader
compare to our new generalized master equation approach and also to
use later to compare the results of the two methods. The starting
point is the following quantum mechanical Hamiltonian 
\begin{align}
H & =\hbar\omega_{c}a^{\dagger}a+\hbar\omega_{m}b^{\dagger}b\\
 & +\hbar g\,a^{\dagger}a\left(b+b^{\dagger}\right)+i\hbar\Omega\left(ae^{i\omega_{L}t}-a^{\dagger}e^{i\omega_{L}t}\right),\nonumber 
\end{align}
where $a$ and $a^{\dagger}$ are the plasmonic ladder operators that
follow the Bosonic commutation relation $\left[a,a^{\dagger}\right]=1$.
Also, $g$ is the coupling factor between the single mode plasmonic
system and the molecule, and $\Omega$ is the pump parameter for the
plasmonic system. This Hamiltonian can be summed over multiple resonances
when necessary, however the numerical/analytical implementations can
be cumbersome. More importantly, it is not obvious how one would adopt
such an approach to a continuum of modes such as in waveguides.

The corresponding master equation to be solved for system dynamics
is 
\begin{align}
\frac{d\rho}{dt} & =-\frac{i}{\hbar}\left[H,\rho\right]\\
 & +\gamma_{c}\left(2a\rho a^{\dagger}-a^{\dagger}a\rho-\rho a^{\dagger}a\right)\nonumber \\
 & +\gamma_{m}\left(\bar{n}^{{\rm th}}+1\right)\left(2b\rho b^{\dagger}-b^{\dagger}b\rho-\rho b^{\dagger}b\right)\nonumber \\
 & +\gamma_{m}\bar{n}^{{\rm th}}\left(2b^{\dagger}\rho b-bb^{\dagger}\rho-\rho bb^{\dagger}\right).\nonumber 
\end{align}
One then calculates the cavity emitted spectrum using 
\begin{equation}
S\left(\omega\right)=\omega^{4}\,{\rm Re}\left[\int_{0}^{\infty}d\tau\,e^{i\left(\omega-\omega_{L}\right)\tau}\left<a^{\dagger}\left(t\right)a\left(t+\tau\right)\right>\right],
\end{equation}
where the quantum regression theorem is employed to find the expectation
value of the two-time correlation function $\left<a^{\dagger}\left(t\right)a\left(t+\tau\right)\right>$.
In general, numerical implementations of this problems are used to
obtain the system spectrum. However, if the so-called linearization
procedure is used \cite{RevModPhys.86.1391,QEDpaperRaman} such that
the evolution of the ladder operators follow a set of linearly coupled
equations, then one can tackle the problem analytically using Laplace
transform. The later approach is what we do in the results section
to compare with our generalized master equation technique for the
plasmonic reservoir. Thus is a fair comparison, as the classical pump
limit that we have employed in our theory is essentially equivalent
to the linearized Hamiltonian \cite{QEDpaperRaman}. Note, also, that
the spectrum defined in this approach does not include the full propagation
effects and the additional $\omega^{4}$ is used to accommodate a
simple dipole propagation argument to the far field. In our approach,
the details of radiation propagation are fully included.

\subsection{QNM expansion of the Green function\label{subsec:QNM-expansion-of}}

In this subsection, we briefly discuss the QNM expansion technique
for the Green function that can be used in a general open cavity systems,
even for plasmonic resonators. While this provides a powerful tool
in analyzing the system response in plasmonic resonators in an efficient
manner, it can also used to simplify our analytically derived SERS
spectrum and, e.g., reveal the limits of the well-known $E^{4}$ enhancement
factor. Extension of the arguments made below to other plasmonic structures
such as waveguides and slabs is straightforward, when the appropriate
alternative Green function expansion is used.

Quasinormal modes, $\mathbf{\tilde{f}}_{\mu}\left(\mathbf{r}\right)$,
are solutions to a non-Hermitian Maxwell's problem subjected to open
boundary conditions. Consequently, they are described by complex frequencies
$\tilde{\omega}_{\mu}=\omega_{\mu}-i\omega_{\mu}/2Q_{\mu}$, where
the imaginary part of which is a measure of losses involved, which
can be due to leakage of energy from the resonator as well as metallic
Ohmic losses. The analytical system Green function can be expanded
in terms of these QNMs through \cite{Leung1994,KristensenACS} 
\begin{equation}
\mathbf{G}^{{\rm QNM}}\left(\mathbf{r}_{1},\mathbf{r}_{2};\omega\right)=\sum_{\mu}A\left(\omega\right)\mathbf{\tilde{f}}_{\mu}\left(\mathbf{r}_{1}\right)\mathbf{\tilde{f}}_{\mu}\left(\mathbf{r}_{2}\right),\label{eq:GF}
\end{equation}
where $A\left(\omega\right)=\omega^{2}/2\tilde{\omega}_{\mu}\left(\tilde{\omega}_{\mu}-\omega\right)$
and $\mathbf{\tilde{f}}_{\mu}\left(\mathbf{r}\right)$ are normalized.
The Green function enables us to connect to a wide range of physical
quantities such as the spontaneous emission enhancement and non-radiative
decay rates from quantum emitters. Normalization of the QNMs can be
done in different ways \cite{Kristensen2012,Sauvan2013}. Note that
the Green function expansion in here is different from standard expansions
seen in Hermitian theories along with the fact that a complex pole
is involved in the QNM expansion.

\begin{figure}
\includegraphics[width=1\columnwidth]{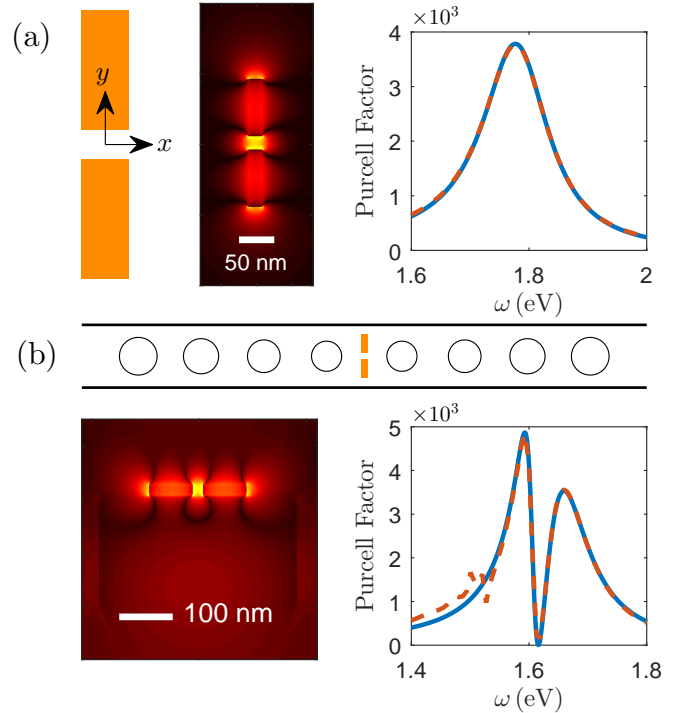}

\caption{(a) Two cylindrical nanorods made of gold are used to form a single
mode plasmonic resonator. The three sub-figures show the dimer schematic,
projected on $xy$ plane, the spatial map of the dimer QNM, also on
the $xy$ plane, and the projected Purcell enhancement along the y
axis at exactly the middle of the dimer gap. (b) The previous dimer
is placed on top of a photonic crystal nanobeam cavity and 5 nm away
from the surface of the beam to form a hybrid system. The three sub-figures
show the top view of the hybrid device, projected to $xy$ plane,
the spatial map of one of the system QNMs, on the $yz$ plane, and
the projected Purcell enhancement along the y axis at exactly the
middle of the dimer gap. Here, blue-solid is calculated using the
QNM expansion of the Green function where the red-dashed is the full
dipole calculations.\label{fig:qnm}}
\end{figure}

As an example resonator system, we first consider a gold dimer where
each nanorod is made of a cylinder with radius of $r_{r}=10$ nm and
height of $h_{r}=80$ nm; see Fig.~\ref{fig:qnm}.(a). The dimer
is placed in free space and the Drude model is used to model its dispersive
and lossy behavior through 
\begin{equation}
\varepsilon(\omega)=1-\frac{\omega_{p}^{2}}{\omega(\omega+i\gamma_{p})},
\end{equation}
where the plasmon frequency and the collision rate are $\omega_{p}=8.29\,{\rm eV}$
and $\gamma_{p}=0.09\,{\rm eV}$, respectively. In the same figure
we have also plotted the spatial map of the QNM supported by the system
where a hot spot is formed in between the 20 nm gap of the dimer.
In addition, we have also plotted the corresponding LDOS at exactly
the center of the dimer gap and projected along the $y$-axis, which
shows a single mode behavior over a wide range of frequencies centered
at $\omega_{c}=1.78$ eV with the quality factor of $Q=13$. The details
of the QNM calculation are presented in Ref.~\cite{hybrid}. 

Next, we place the exact same dimer on top of a photonic crystal nanobeam
cavity to create a hybrid photonic-plasmonic system with multimode
characteristics, as schematically shown in Fig.~\ref{fig:qnm}.(b).
The nanobeam PC-cavity, previously studied in \cite{hybrid}, is modeled
as silicon-nitride with refractive index of $n=2.04$, where the height
of $h_{b}=200$ nm, and the width of $W_{b}=367$ nm. This design
uses a mirror section as well as a taper section, the details of which
are given in \cite{hybrid}, to obtain a large quality factor of $Q=3\times10^{5}$
at the resonance frequency of $\omega_{c}=1.62$ eV. Due to the coupling
between the dimer and PC-cavity, very strong hybridization of the
individual mode takes place. The resonance frequencies of the two
hybridized QNMs are found to be $\omega_{1}=1.64\,{\rm eV}$ and $\omega_{2}=1.61\,{\rm eV}$,
with the corresponding quality factors of $Q_{1}=15$ and $Q_{2}=55$,
respectively. In Fig.~\ref{fig:qnm}.(b), we plot the spatial map
of the first QNM with a lower $Q$, where a clear signature of hybridization
of the individual modes is seen; the same feature was also seen for
the other QNM with the higher $Q$. When both of the hybrid system
QNMs are used, an accurate representation of the system LDOS is obtained
over a wide range of frequencies as shown in the same figure. The
most important feature of the hybrid LDOS is the interference between
the two main QNMs that results in a strong modification of the spontaneous
emission rate of a quantum emitter at that place, such that there
will be minimum that the emission is significantly reduced (to around
5) \cite{hybrid}.

We can now use the QNM representation of the system Green function
discussed above to extract the $E^{4}$ dependence of our analytical
expression for the SERS spectrum that is always present in both of
the pump power regimes discussed before. The first key element in
Eq.~\eqref{eq:spectrum2} is the enhancement at pump frequency through
$\eta$. Without loss of generality, we consider the Green function
expansion of Eq.~\eqref{eq:GF} when only one QNM is present, namely
$\mathbf{\tilde{f}}\left(\mathbf{r}\right)$. Using this approximation,
and assuming that the initial pump field is launched along $\mathbf{n}$,
$\eta$ can be written as 
\begin{equation}
\eta=A\left(\omega_{L}\right)\left[\mathbf{n}\cdot\mathbf{\tilde{f}}\left(\mathbf{r}_{m}\right)\right]\left\{ \int\left[\varepsilon\left(\mathbf{r},\omega_{L}\right)-\varepsilon_{B}\right]\mathbf{\tilde{f}}\left(\mathbf{r}\right)d{\bf r}\right\} ,
\end{equation}
where $A\left(\omega_{L}\right)$ can be thought the resonator quality
factor, $Q$, especially when on resonance, and the spatial integration
incorporates a spatial integration over the mode. Importantly, when
used in Eq.~\eqref{eq:spectrum2}, the modal field value at the molecule
location, $\mathbf{\tilde{f}}\left(\mathbf{r}_{m}\right)$, recovers
the squared dependence of the SERS spectrum at the laser frequency.

Next, we consider the propagation effect through $\left|\mathbf{G}\left(\mathbf{R},\mathbf{r}_{{\rm m}};\omega\right)\cdot\mathbf{n}\right|^{2}$.
This can be expanded using the system Green function, to find 
\begin{equation}
\mathbf{G}\left(\mathbf{R},\mathbf{r}_{{\rm m}};\omega\right)\cdot\mathbf{n}=A\left(\omega\right)\mathbf{\tilde{f}}\left(\mathbf{R}\right)\left[\mathbf{\tilde{f}}\left(\mathbf{r}_{m}\right)\cdot\mathbf{n}\right],
\end{equation}
where $\mathbf{\tilde{f}}\left(\mathbf{R}\right)$ is the modal value
at the point detector, and $\mathbf{\tilde{f}}\left(\mathbf{r}_{m}\right)\cdot\mathbf{n}$
is the modal value at the molecule location that appeared before inside
$\eta$. Taking everything into account, one recovers the well-known
power of four rule, $|\mathbf{\tilde{f}}\left(\mathbf{r}_{m}\right)|^{4}$,
for the detected SERS spectrum.

Similar consideration can be easily made for the imaginary part of
the Green function at the molecule location, ${\rm Im}\left\{ G_{nn}\left(\mathbf{r}_{{\rm m}},\mathbf{r}_{{\rm m}};\omega\right)\right\} $,
to find that $J_{{\rm ph}}$ defined in Eq.~\eqref{eq:defineJph}
is also proportional to $|\mathbf{\tilde{f}}\left(\mathbf{r}_{m}\right)|^{4}$.
Therefore, at the high pump powers discussed before, we now find that
Raman signal enhances as $|\mathbf{\tilde{f}}\left(\mathbf{r}_{m}\right)|^{8}$
instead of $|\mathbf{\tilde{f}}\left(\mathbf{r}_{m}\right)|^{4}$.

\section{Results\label{sec:Results}}

We now present some example results of ourSERS theory using the plasmonic
environments discussed previously. Our analysis will include the role
of propagation on the detected Stokes and anti-Stokes signals for
different pump scenarios, the effect of pump detuning from the plasmonic
resonances on the detectable intensities, as well computing the spatial
map of the Stokes (anti-Stokes) intensity profiles. For the molecular
vibrations, we use the particular mode of the R6G molecule at $\omega_{m}=0.16$
eV \cite{R6G} where the thermal dissipation rate is taken to be $\gamma_{m}=1.6$
meV.

\begin{figure}
\includegraphics[width=1\columnwidth]{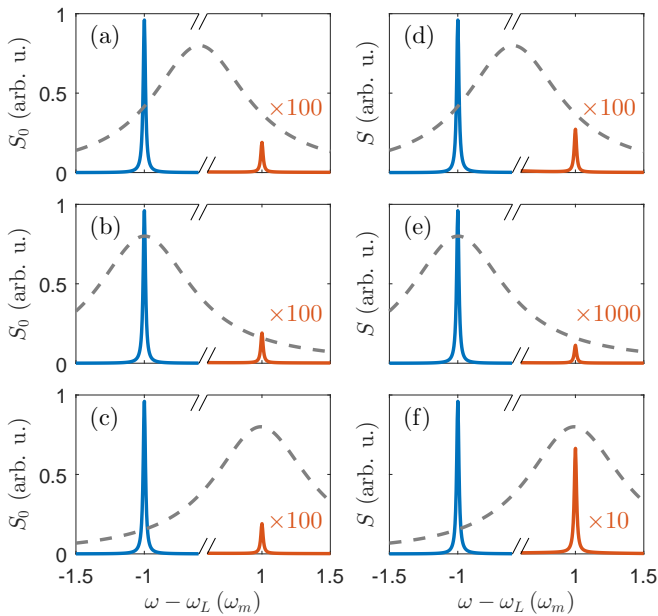} \caption{Raman spectra calculated for the R6G molecule coupled to the gold
dimer, using Eq.~(\ref{eq:spectrum2}) for three different laser
pump frequencies: (a)/(d) where $\omega_{L}=\omega_{c}$, (b)/(e)
where $\omega_{L}=\omega_{c}+\omega_{m}$ and (c)/(f) where $\omega_{L}=\omega_{c}-\omega_{m}$.
The plots on the left only show the $S_{0}\left(\omega\right)$ (emitted
spectrum), where in contrast, on the right, the full far-field (detected)
spectrum is plotted. The same pump power, $\varepsilon_{0}c\left|\mathbf{F}_{0}\right|^{2}/2=130\,{\rm mW/\mu m^{2}}$,
and the temperature, $T=300\,{\rm K}$, is assumed in all calculations.
The gray-dashed line on the background of all plots indicates the
LDOS profile of the plasmonic resonance, and the magnifying factors
are only applied to the anti-Stokes intensities.\label{fig:propagation}}
\end{figure}

We begin by discussing the propagation affects on the detected spectrum.
The spectrum calculated using Eqs.~\eqref{eq:spectrum2}, \eqref{eq:spectrum3}
and \eqref{eq:spectrum4} is plotted in Fig. \ref{fig:propagation}
for three different excitation scenarios, where in each case, both
the emitted spectrum quantified using $S_{0}\left(\omega\right)$
(left column) and the detected spectrum\textemdash quantified using
$S\left(\mathbf{R},\omega\right)$ when the detector placed $x=500\,{\rm nm}$
away form the dimer (right column)\textemdash are presented (the results
are qualitatively the same for larger distances). The three scenarios
are as follows: (a)/(d) $\omega_{L}=\omega_{c}$, where the pump laser
is on-resonance with the plasmonic mode, (b)/(e) $\omega_{L}=\omega_{c}+\omega_{m}$,
where the laser pump is detuned to the blue side of the plasmonic
resonance by exactly the frequency of R6G vibration, and finally (c)/(f)
$\omega_{L}=\omega_{c}-\omega_{m}$, where the laser pump is detuned
to the red side of the plasmonic resonance by exactly the frequency
of R6G vibration. All the calculation assume room temperature where
$n^{{\rm th}}=0.002$, and the same fixed value for the pump intensity.
As discussed in the theory section, the spectral properties of the
Raman signals at the generation stage can be quite different to those
that are detected. In particular, the difference depends on the operating
frequency for the pump field and the changes of the LDOS between the
laser drive frequency and at the Raman sidebands. For example, by
comparing Fig.~\ref{fig:propagation}.(a) and Fig.~\ref{fig:propagation}(d),
where the system is pumped on resonance with the plasmonic mode, the
propagating effect to the far-field seems to be a bit in favor of
more enhancement for the anti-Stokes signal than the Stokes signal.
This difference becomes even more pronounced when one excites the
system in a red detuning configuration where the anti-Stokes emission
exploits the maximum enhancement form the plasmonic environment through
propagation. Indeed, in the far-field the anti-Stokes can gain a higher
value than the Stokes intensity under high pump intensity that are
not shown.

\begin{figure}
\includegraphics[width=1\columnwidth]{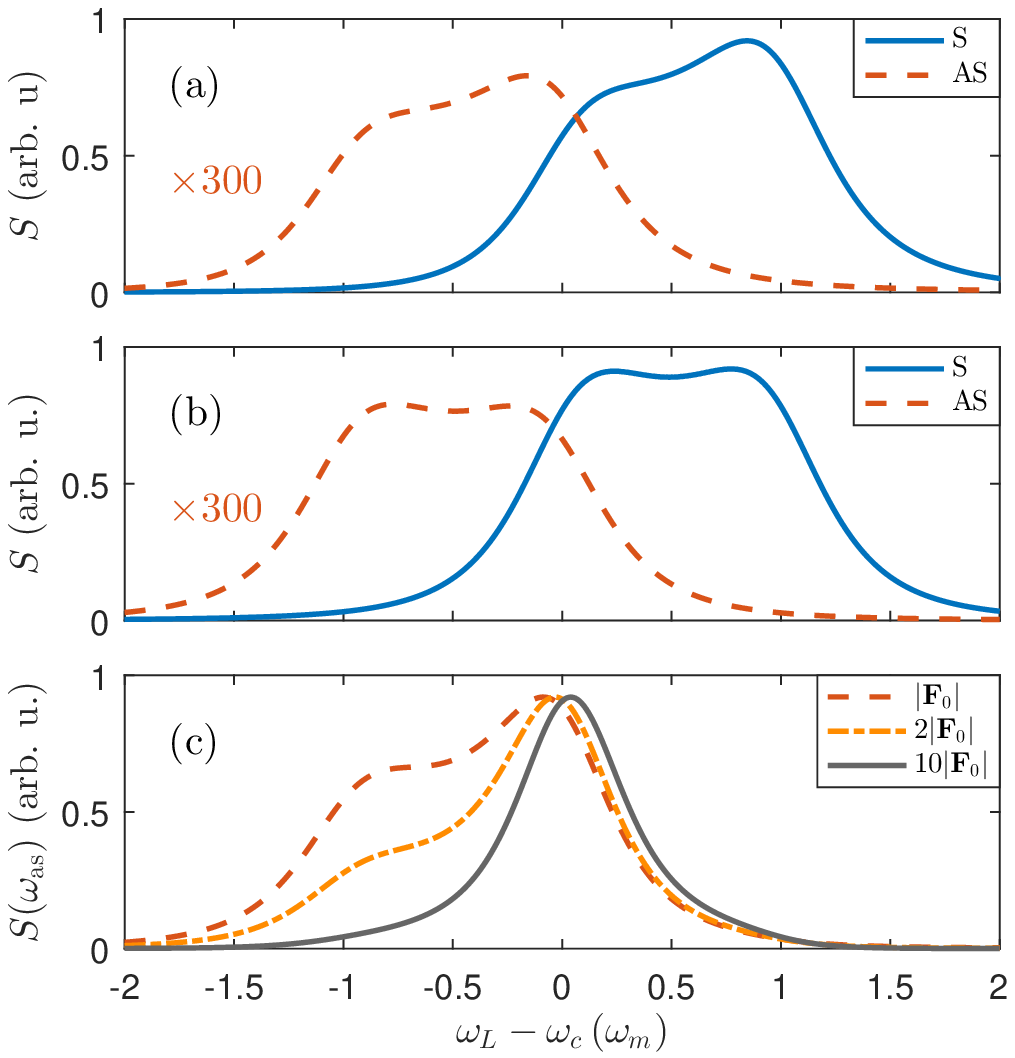}

\caption{Plot of Stokes and anti-Stokes intensity peaks calculated as a function
of laser detuning when: (a) The plasmonic dimer is assumed to be having
a Lorentzian line shape and the quantum optomechanical model of subsection
\ref{subsec:Existing-quantum-optomechanical} is used where the pump
power is set to $\Omega=1\,{\rm eV}$. (b) The true gold dimer response
is used through inclusion of the system Green function in Eq.~\eqref{eq:spectrum2}
and the pump power is set to $\varepsilon_{0}c\left|\mathbf{F}_{0}\right|^{2}/2=130\,{\rm mW/\mu m^{2}}$
where using the recipe in Ref.~\cite{QEDpaperRaman} translates to
approximately $\Omega=1\,{\rm eV}$. (c) Shows the effects of increasing
the pump intensity on the anti-Stokes detectable intensities when
going from red-dashed to gray-solid using our generalized Master equation
technique. The Stokes signals were quite similar when the pump intensity
was increased and the starting pump power is set to $\varepsilon_{0}c\left|\mathbf{F}_{0}\right|^{2}/2=130\,{\rm kW/\mu m^{2}}$.\label{fig:detuned}}
\end{figure}

Next, we consider the effect of laser detuning from the plasmonic
resonance on the Stokes and anti-Stokes intensities. Different cases
are considered in Fig.~\ref{fig:detuned}, where the Stokes and anti-Stokes
signal are calculated as follows: (a) using the quantum optomechanical
description briefly discussed in subsection \ref{subsec:Existing-quantum-optomechanical}
for the R6G molecule coupled to the gold dimer, (b) using our analytical
expression for the detectable spectrum given in Eq.~\eqref{eq:spectrum2}
where the gold dimer Green function is accurately calculated using
the QNM expansion of Eq.~\eqref{eq:GF}. As seen, the agreement between
the two methods is qualitatively good, since the response of the system
happens to be very similar to a Lorentzian lineshape, but there are
still quantitative differences. Based on our investigations, we attribute
these differences to our estimation of the pump enhancement effect
which involves calculating $\eta$ through use of Eq.~\eqref{eq:eta},
which in turn employs the full system Green function. To make this
clearer, we note an excellent agreement between the prediction of
our model and the modal quantum optomechanical model of Ref.~\cite{QEDpaperRaman},
if $\eta$ was estimated using the plasmonic LDOS (which is implied
in the theory we compare with) rather than using Eq.~\eqref{eq:eta}.

In addition, we further consider three different pump intensity scenarios
in Fig.~\ref{fig:detuned}(c). As seen, for the the R6G molecule
coupled to the gold dimer, increasing the pump field reshapes the
anti-Stokes signal such that a single peak feature at a different
location is obtained. For the same pump values, small changes to the
general shape of the Stokes signal was noticed. A similar finding
was made in \cite{QEDpaperRaman} suggesting that increasing the pump
intensity moves the anti-Stokes signal peak toward the Stokes signal
peak in a Raman experiment, due to the nonlinear interactions. Even
though this effect is attributed to the nonlinear behavior under intense
pumps, note that the $J_{{\rm ph}}$ terms in our master equations
are proportional to square of the pump field in comparison to the
thermal dissipation terms, and it is important to appreciate the fact
that the Raman spectrum is being generated differently than to what
is measured in a plasmonic environment. Therefore, the counter intuitive
enhancement mechanism argued before is what leads to the generally
expected SERS spectrum, where the detuning dependencies of Raman signals
for different pump values are correctly captured. Indeed, the results
argued earlier would have not been obtained if one used an alternative
argument. It is also worth noting that the pump values used in Fig.~\eqref{fig:detuned}
are such that $\varepsilon_{0}c\left|\mathbf{F}_{0}\right|^{2}/2=1.3\times10^{5}\,{\rm W/\mu m^{2}}$
which is extremely high. This is particularly due to lower plasmonic
enhancement achieved in using the gold dimer, e.g., in comparison
to what used in \cite{QEDpaperRaman}, where a mode volume orders
of magnitude smaller than what we have used here is considered. Accordingly,
our estimation of the coupling factor $g$ used in the optomechanical
Hamiltonian of Sec. \ref{subsec:Existing-quantum-optomechanical}
for our gold dimer is five orders of magnitude smaller than what used
in \cite{QEDpaperRaman}.

\begin{figure}[h]
\includegraphics[width=1\columnwidth]{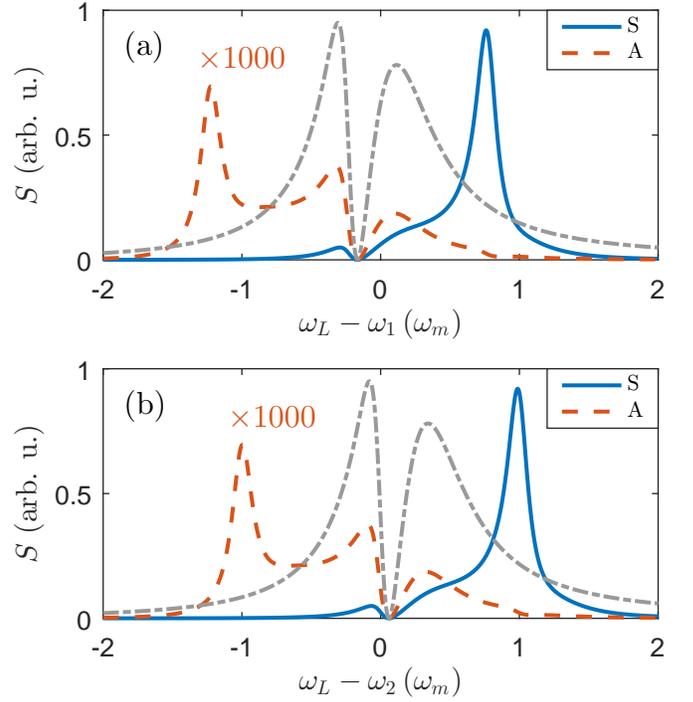} \caption{Stokes (blue-solid) and anti-Stokes (red-dashed) detection intensities
for the R6G molecule when coupled to the hybrid device at exactly
the middle of the dimer gap and oriented along the y axis. The plasmonic
LDOS is projected in gray in the background for comparison. (a) and
(b) show representation of the same data for when the laser detuning
is measured with respect to $\omega_{1}$ and $\omega_{2}$, respectively.\label{fig:hybrid}}
\end{figure}

Next we demonstrate that our theory can easily be applied to more
complex plasmonic environments where the simple single mode (Lorentzian-like)
picture does not work. Such a task is not easy using a coupled mode
quantum theory for computing the SERS, even if several modes were
employed. For example, even ignoring the conceptual difficulties of
treating the plasmon mode as a normal mode, numerical quantum mechanical
calculation of the SERS when there are three modes involved, one for
the molecule and two for the resonator, can be numerically cumbersome,
since typically the size of the truncated Hilbert space depends strongly
on the number of modes involved (each one quantized) and therefore
obtaining converged are highly computationally demanding. In contrast,
our theory in its existing form is capable of overcoming such challenges.

As an example non-single mode system, we consider the hybrid photonic-plasmonic
structure discussed in Sec.~\ref{subsec:QNM-expansion-of}. Because
the hybrid structure supports more than one QNMs, the resulting spectrum
can be naturally different. In particular, one might wonder if there
are certain aspects of the Raman process that is derived by one particular
QNM. Using the analytical results or our theory, Eq.~\eqref{eq:spectrum2},
and the QNM representation of the system Green function for the hybrid
system through expansion of Eq.~\eqref{eq:GF}, we plot the Stokes
(blue-solid) and anti-Stokes (red-dashed) detected intensities for
different pump detuning in Fig.~\eqref{fig:hybrid}. These results
are displayed in two different ways: (a) the detuning is with respect
to the first hybrid resonance at $\omega_{1}=1.64$ eV that is associated
with the lower $Q$ QNM, and is confirmed to be more dimer-like; (b)
detuning is quantified with respect to the second hybrid resonance
at $\omega_{2}=1.61$ eV, that is associated with the higher $Q$
QNM, and is confirmed to be less dimer-like. We have also projected
the hybrid LDOS on the background in gray dotted-dashed for guidance.
As seen, when the LDOS is suppressed to near zero due to the interference
between the QNMs, the Raman intensities follow the same trend. In
addition, there seems to be three high intensity operation points
available for both Stokes and anti-Stokes detections. More importantly,
the maximum detection seems to be one $\omega_{m}$ to the right (left)
for the Stokes (anti-Stokes) in Fig.~\eqref{fig:hybrid}(b) where
the detuning is being measured with respect to the $\omega_{2}$.
Therefore, in this specific configuration of the hybrid system, the
SERS signals seem to be coupled to the QNM frequency $\omega_{2}$
that is primarily attributed to the nanobeam cavity resonance rather
than the gold dimer. This is not a general conclusion to be made and
for different configurations of the hybrid device, depending on the
regime where the two QNMs are coupled \cite{hybrid}, other outcomes
might be feasible. 

\begin{figure}[t]
\includegraphics[width=1\columnwidth]{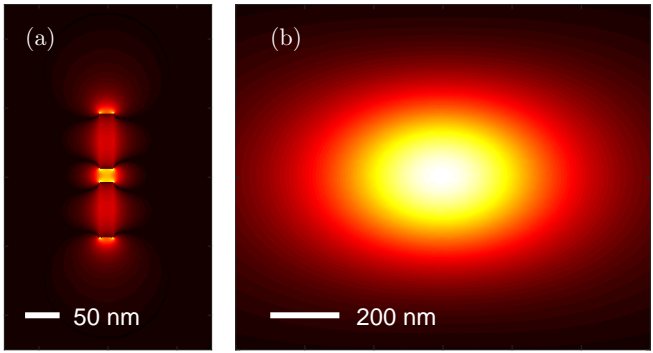} \caption{Spatial map of the detected Stokes intensity from the R6G molecule
coupled to the Gold dimer, (a) when the molecule is moved around the
and the detector is fixed at 500 nm away from the dimer to the right,
(b) when the molecule is fixed at the middle of the dimer gap and
the detector is scanned at a plane parallel to the longitudinal axis
of the dimer and 500 away form it. Nonlinear scaling of the values
are done for visualization purposes.\label{fig:map}}
\end{figure}

Our model also allows accurate and rapid calculation of both Stokes
and anti-Stokes intensities as a function of the spatial location
both for the molecule and for the detector. Although, previous theories
can effectively employ a position dependent coupling factor for the
molecule at different location in order to attempt such scenarios,
this can be numerically inefficient specially when multiple resonances
are involved, for the reasons elaborated before. We will demonstrate
this below for the single mode gold dimer. Assuming that the detection
setup is fixed at a particular location (far from the dimer), one
can move the molecule around the plasmonic resonator and see how both
Stokes and anti-Stokes emissions are dependent upon that. In Fig.~\ref{fig:map}.(a),
we have plotted such a map only for Stokes intensity (the profile
is confirmed to be the same for anti-Stokes but having relatively
lower values for the condition under investigation) which confirms
that the Raman emission is dominated by the modal shape of the plasmonic
mode, namely the Purcell factor behavior of a quantum emitter placed
nearby. As seen, in this case, it may be desired for the molecule
to be placed in the middle of the dimer gap in order for the Raman
response to be optimized. In contrast, for a fixed molecule location,
one might want to move the detector around and scan for the maximum
intensities. The detector location dependence of the Stokes signal
is plotted in Fig.~\ref{fig:map}.(b) where a dipole emission in
the far field is seen. As mentioned before, such spatial maps can
be produced as easily for other plasmonic geometries such as waveguides,
without any need to modify our theory. All one needs, is an accurate
representation of the total system Green function that in our case
was made available through use of QNMs.

\section{Conclusions\label{sec:Conclusions}}

We have developed a powerful and intuitive Green function based formalism
that can be used to model SERS using plasmonic systems of arbitrary
geometry. For example, the plasmonic environment can be either a resonator,
a waveguide, or even a more complex structure such as a hybrid photonic-plasmonic
platform. A generalized master equation is employed such that the
plasmonic environment is treated as a system bath and the system dynamics
are projected onto the basis of the molecular vibrations. Therefore,
while the plasmonic degrees of freedom are not present, the exact
system Green function is self-consistently incorporated to account
for true plasmonic characteristic such as emission enhancement and
quenching. As a result, analytical expressions in the most general
form are given for the SERS spectrum near plasmonic environments that
can be used to quantify Stokes and anti-Stokes emission intensities.
We reported that the dynamics of the SERS is quite different in generation
stage than in propagation stage. The induced Raman polarization seems
to be emitting photons at Stokes frequency proportional to the plasmonic
enhancement at the co-existing anti-Stokes frequency and vice-versa.
However, plasmonic LDOS comes into play differently at propagation
stage such that both Stokes and anti-Stokes photons are enhanced at
their respective frequencies. In comparison to the excising models
mostly working in a single mode regime, our model simply extends beyond
this limitation and characterizes the Raman spectrum as a function
of detuning, pump power, spatial location of the molecule/detector
etc. Given the analytical nature of our results, it can be used to
model any SERS experiment in no time, once the system Green function
is accurately available.

\section*{Acknowledgment}

We thank Queen's University and the Natural Sciences and Engineering
Research Council of Canada for financial support. We would also like
to thank Amr Helmy, Herman M. K. Wong and Reuven Gordon for useful
discussions.

\bibliographystyle{apsrev4-1}
\bibliography{Refs}

\end{document}